\documentclass[12pt]{article}
\usepackage[a4paper, total={6in, 8in}]{geometry}
\newcommand{\dd}{\mathop{}\!\mathrm{d}}
\usepackage{amsmath}
\usepackage{mathtools}
\newcommand{\matr}[1]{#1}  

\usepackage{authblk}
\usepackage{hyperref}
\date{}	
\title{Spatial Resolution Enhancement of Oversampled Images Using Regression Decomposition and Synthesis}
\author[a]{Hsien-Wei Chen\thanks{email: hw.chen2018@gmail.com}}
\affil[a]{Department of Research and Development, Precima, Nielsen Global Connect, Toronto, Ontario, Canada}

\begin{document}

	\begin{titlepage}
	\maketitle\thispagestyle{empty}
	\begin{abstract}
	A new statistical model designed for regression analysis with a sparse design matrix is proposed. This new model utilizes the positions of the limited non-zero elements in the design matrix to decompose the regression model into sub-regression models. Statistical inferences are further made on the values of these limited non-zero elements to provide a reference for synthesizing these sub-regression models. With this concept of the regression decomposition and synthesis, the information on the structure of the design matrix can be incorporated into the regression analysis to provide a more reliable estimation. The proposed model is then applied to resolve the spatial resolution enhancement problem for spatially oversampled images. To systematically evaluate the performance of the proposed model in enhancing the spatial resolution, the proposed approach is applied to the oversampled images that are reproduced via random field simulations. These application results based on different generated scenarios then conclude the effectiveness and the feasibility of the proposed approach in enhancing the spatial resolution of spatially oversampled images.
	\end{abstract}
	

	\textbf{Highlights}
	\begin{itemize}
		\item A new statistical model for regression analysis with a sparse design matrix
		\item A solution to the enhancement of the spatial resolution of oversampled images
		\item An algorithm to provide the flexibility in sensor designs
	\end{itemize}
	\end{titlepage}

	\section{Introduction}
	Via continuously scanning the Earth surface, remote sensing images provide regional observations for monitoring environmental parameters. However, to ensure that these regional observations depict the details of the scanned areas, the unit area that the sensor scans needs to be small enough to describe the textures of the objects of interest. In other words, depending on the application, different levels of the spatial resolution of a remote sensing image are required to provide useful information. For retrieving information on sea surface temperatures, passive microwave images with a spatial resolution of tens of kilometers are acceptable (O'Carroll et al., 2008; Reynolds et al., 2007). For monitoring land covers, optical remote sensing images with a spatial resolution of hundreds of meters are required (Hall et al., 2002; Strahler et al., 1999; Zhang et al., 2003). For numerically simulating floods in urban areas, digital elevation models with a spatial resolution of tens of meters are not sufficient for providing the required information on buildings (Haile and Rientjes, 2005; Schubert and Sanders, 2012). Consequently, the mechanism to control spatial resolution is the key to the applications of remote sensing images.
	
	For this purpose, controlling the distance between the sensor and the Earth surface is the most direct approach. However, this direct approach is difficult to implement in practice since choosing the platform that mounts the sensor to adjust the spatial resolution of remote sensing images for a specific application is unrealistic. Instead, the speed of scanning the Earth surface can be designed to reduce the distance between adjacent areas to be scanned, to offset the deficit due to insufficient spatial resolution. However, the direct usage of the images obtained from a sampling frequency that is higher than the required minimum frequency to scan continuously the Earth surface without gaps, still provides limited improvements in spatial resolution. The main reason is that the unit area that the sensor scans is not broken down to a smaller unit via increasing the sampling frequency. In other words, to effectively use this additional information that comes from the high scanning frequency for enhancing spatial resolution, a post-process procedure is required.
	
	With the information on the positions of the sensor relative to the Earth surface, the viewing directions of the sensor, and the visual fields of the sensor, the relationship between the observed records collected by the sensor and the to-be-converted image with the desired spatial resolution can be formulated. However, since the spatial resolution enhancement problem is essentially a disaggregation problem that aims to spatially decompose each signal observed by the sensor, this formulated relationship based on the properties of the sensor cannot guarantee a unique conversion from the observed records of the sensor to the spatial-resolution-enhanced image. In order to identify an appropriate conversion for enhancing the spatial resolution, additional information that comes from temporally related images (Yu et al., 2006) and spatially related images (Aiazzi et al., 2002; Gross and Schott, 1996; Hu et al., 2019) can be used as a reference. By using the matrix decomposition techniques, the matrix that describes the relationship between the observed records collected by the sensor and the to-be-converted image with the desired spatial resolution, can be modified to obtain a unique conversion (Arai and Matsumoto, 1994; Wiman, 1992). By applying constrained optimization, a corresponding conversion can be found via controlling the variation of the pixel values in the to-be-converted image (Farrar and Smith, 1992; Yanovsky et al., 2014). However, these image conversion selection approaches do not fully utilize the spatial information collected via the increased sampling frequency, because the information on the distances between the pixels in the to-be-converted image and the locations on the Earth surface to which the observations collected by the sensor correspond, is not taken into consideration.

	To utilize this distance information for enhancing the spatial resolution, the concept of the regression decomposition and synthesis is proposed. By using this distance information to decompose regression models, the appropriate conversions from the observed records of the sensor to the spatial-resolution-enhanced image can be identified. By synthesizing these identified conversions, this distance information can further be incorporated as weights to reinforce the reliability of the spatial-resolution-enhanced image.
	
	\section{Methodology}
	To ensure that the signals received by a sensor are stronger than the corresponding noises when sensing, the sensor should have an instantaneous field of view (IFOV) that can collect enough radiance. This IFOV and the corresponding sensing distance then limit the spatial resolution of the sensed image. In order to retrieve more information for practical needs, an alternative solution is to allow oversampling to have overlaps between the instantaneous footprints of the sensor. However, the resulting image is blurred with limited information that can be read when displaying. To resolve this problem, an appropriate mesh grid can first be defined on the Earth surface. Then, the radiance value for each grid can be solved via the linear equation set that represents the at-sensor radiances as the weighted sums of the values of these non-overlapped grids. With this approach, the information that comes from oversampling can be fully utilized.
	
	To solve this linear equation set is essentially to find the weighted sum of all the at-sensor radiance observations that represents the radiance value of each grid. However, when the total area on the Earth surface scanned by the sensor to form the sensed image is large, this weighted-sum estimation does not make sense. In this situation, most of the at-sensor radiance observations used to estimate the radiance value for each grid does not contain any radiance information on the corresponding grid. The reason for the inclusion of the redundant at-sensor radiance observations in this weighted-sum estimation is because of the overlaps between the instantaneous footprints of the sensor. Due to the overlaps, estimating the radiance value of a grid requires the radiance values of the grids next to it. Estimating these adjacent grids again relies on the grids next to them. Eventually, all the grids are tied together in one estimation. Estimating the radiance value of each grid needs all the at-sensor radiance observations. Nonetheless, for each grid, the inclusion of these redundant at-sensor radiance observations has little to no contribution to the weighted-sum estimation. Often, the weighted-sum estimation becomes unreliable or unsolvable due to these redundant observations. To ensure that the solution to each grid is reliable, the redundant observations need to be excluded from the linear equation set to define a {\it local} scope for the weighted-sum estimation. The accuracy of the resulting weighted-sum estimation needs to be evaluated as the reference for the reconstruction of the optimal estimation in the {\it global} scope.
	
	{\it Radiance signal}. Based on the definition of radiance, the radiance signal received by the sensor can be represented as the total radiant energy collected by the sensor divided by the IFOV of the sensor, the equivalent exposure area of the sensor surface, and the exposure time of the sensor. As a consequence, the radiance \(L^{s}_{p,\omega}\) received by the sensor \(s\) at the position \(p\) with the viewing direction \(\omega\) can be expressed as an integral of the radiance \(L_{h}\) emitted by the point \(h\) on the Earth surface, over the corresponding constant, as illustrated in Equation (1).
	
	\begin{equation}
	L^{s}_{p,\omega} = \frac{1}{\iint \limits_{f,h} \cos \theta_{f,h} \dd A_{f} \dd \Omega_{f,h}} \iint \limits_{f,h} L_{h} \cos \theta_{f,h} \dd A_{f} \dd \Omega_{f,h}
	\end{equation}
	
	where \(f\) and \(\dd A_{f}\) represent a point on the sensor surface and the corresponding area of the point \(f\), respectively, \(h\) and \(\dd \Omega_{f,h}\) are a point on the Earth surface and the corresponding solid angle of the field of view from \(f\) that \(h\) covers, respectively, and \(\theta_{f,h}\) represents the angle between the normal vector of the area \(\dd A_{f}\) and the vector from \(f\) to \(h\). The cosine of \(\theta_{f,h}\) is the ratio of the exposure area at \(f\) that receives the radiance from \(h\) over the area \(\dd A_{f}\). The double integrals represent the integration over all the points \(f\) on the sensor surface. For each point \(f\) on the sensor surface, this integration is over all the points \(h\) on the Earth surface that are viewable by this point \(f\) when the sensor \(s\) is at position \(p\) with the viewing direction \(\omega\). Essentially, the numerator and the denominator are the total radiant energy received by the sensor per unit time and the corresponding factor that converts the unit of the numerator to the radiance unit, respectively.
	
	The at-sensor radiance expression illustrated in Equation (1) is derived based on the assumption that the radiance emitted by the Earth surface is uniform in all directions and remains the same when received by the sensor. The information on the atmosphere and terrain can be added to Equation (1) to describe the difference between the radiance emitted by the Earth surface and the radiance received by the sensor to allow an at-sensor radiance expression closer to the reality. However, as the atmospheric and the topographic corrections for remote sensing images are not the main focus in this study, Equation (1) is an appropriate at-sensor radiance model to continue the discussions.
	
	If a mesh grid is defined on the Earth surface, and \(L_{h}\) represents the equivalent radiance emitted by the grid with the center \(h\), then Equation (1) can be approximated by and expressed as the weighted summation that sums over all \(h\) that are viewable by the sensor \(s\) at the position \(p\) with the viewing direction \(\omega\). All the weights of this summation are positive and are summed to one. The weight to which the grid center \(h\) corresponds is proportional to the integration of \(\Delta \Omega_{f,h} \cos \theta_{f,h} \dd A_{f}\) over all the points \(f\) on the sensor, where \(\Delta \Omega_{f,h}\) is the solid angle of the field of view from \(f\) that the grid with the center \(h\) covers. By further considering the uncertainties related to the radiance measurement error, the at-sensor radiance observations can be modeled by Equation (2).
	
	\begin{equation}
	\widetilde{Y} = \matr{X} \beta + \tilde{\epsilon}
	\end{equation}
	
	where \(\widetilde{Y}\) is a column vector with length \(n\) that collects all the at-sensor radiances \(L^{s}_{p,\omega}\) measured by the sensor \(s\) at different positions \(p\) with different viewing directions \(\omega\). \(\beta\) is a column vector with length \(m\) that collects all the equivalent radiances \(L_{h}\) emitted by the grids with different centers \(h\) that are viewable by the sensor \(s\), at least, in one of the position \(p\) and the viewing direction \(\omega\) combinations during sensing. \(\matr{X}\) is a \(n \times m\) matrix in which the weights of the summations that approximate Equation (1) are assigned to the corresponding elements based on the definitions of \(\widetilde{Y}\) and \(\beta\) to write this approximation through linear algebra. Due to the fact that not all the grids defined on the Earth surface are viewable by the sensor at a certain position with the corresponding viewing direction, \(\matr{X}\) is a sparse matrix with few non-zero elements. However, there is no column or row that is full of zeros in \(\matr{X}\) since the elements collected in \(\beta\) are the radiances emitted by the grids that are viewable by the sensor during sensing. Furthermore, because of the weight definition that follows the linear approximation of Equation (1), all the elements in \(\matr{X}\) are greater than or equal to zero, and the sum of all the elements of each row in \(\matr{X}\) is one. \(\tilde{\epsilon}\) is a column vector with length \(n\) in which each element is a random variable that represents the noise of the at-sensor radiance observation in the same row in \(\widetilde{Y}\) during sensing. To trace the source of the noise, each element in \(\tilde{\epsilon}\) is further assumed to be the sum of all the random variables that represent the noises that come from sensing the equivalent radiance emitted by each grid on the Earth surface that is viewable by the sensor during sensing. In other words, for each non-zero element \(x_{i,j}\) in \(\matr{X}\), there is an associated random variable \({\epsilon_{i,j}}\) to represent the corresponding noise. The \(i^{\text{th}}\) element \({\epsilon_{i}}\) in the vector \(\tilde{\epsilon}\) is the sum of the random variable \({\epsilon_{i,j}}\) over all the possible subscripts \(j\). Furthermore, since the random variable \({\epsilon_{i,j}}\) represents the noise, assuming that \({\epsilon_{i,j}}\) are mutually independent with zero mean is reasonable. Otherwise, the sensor system may not be well designed to retrieve information properly. To ensure that the level of uncertainty is reflected by the weights, the standard deviation of the noise \({\epsilon_{i,j}}\) is assumed to be proportional to the non-zero element \(x_{i,j}\) and expressed as \(x_{i,j} \sigma\) via assuming that there is a constant \(\sigma\).
	
	{\it Local regression}. To fully utilize the information that comes from the overlaps between the instantaneous footprints of the sensor to improve the spatial resolution of the sensed image, the size of the grids defined on the Earth surface should be approximately the same as the areas expended by the distances from the center of a instantaneous footprint of the sensor to the center of its adjacent instantaneous footprints of the sensor. In other words, the length \(m\) of the vector \(\beta\), which represents the number of the grids, should be approximately the same as the length \(n\) of the vector \(\widetilde{Y}\), which represents the number of the at-sensor radiance observations. When \(m\) is equal to \(n\), \(\beta\) can be estimated by the inverse of \(\matr{X}\) multiplied by \(\widetilde{Y}\). However, there is no guarantee that \(\matr{X}\) is invertible. Even if \(\matr{X}\) is invertible, this estimation approach for \(\beta\) is unreliable because of the noises that are attached to the at-sensor radiance observations and are propagated to the estimation of \(\beta\) via the inverse of \(\matr{X}\).
	
	The inversion of \(\matr{X}\) provides \(m\) sets of weights to sum over the elements in \(\widetilde{Y}\) to estimate the corresponding \(m\) elements in \(\beta\). However, because of the noises attached to the at-sensor radiance observations, for each element in \(\beta\), not all the elements in \(\widetilde{Y}\) contribute valuable information to the corresponding weighted-sum estimation. To obtain a more reliable weighted-sum estimation for each element in \(\beta\), the elements in \(\widetilde{Y}\) that contribute little information to the estimation of the corresponding element in \(\beta\) should be removed from the estimation of the corresponding element in \(\beta\). To evaluate the usefulness of each element in \(\widetilde{Y}\) when estimating each element in \(\beta\), a direct approach is to use, whether or not the corresponding grid is within the corresponding instantaneous footprint of the sensor, as the criterion. However, since the estimation of each element in \(\beta\) relies on the system of linear equations that characterizes the relationship between the at-sensor radiance observations and the radiances emitted by the grids, independently estimating each element in \(\beta\) is not possible. The removal of the elements in \(\widetilde{Y}\) from the weighted-sum estimations for the elements in \(\beta\) also depends on the criterion that defines the elements in \(\beta\) to be estimated at the same time. 
	
	Since the at-sensor radiance is the unit of the observations, the instantaneous footprints of the sensor would be an appropriate unit to define the local scope to exclude the at-sensor radiance observations that do not contain useful information for estimating the equivalent radiances emitted by the grids defined on the Earth surface. In other words, the \(i^{\text{th}}\) local regression is defined for the \(i^{\text{th}}\) at-sensor radiance observation \(Y_{i}\) in \(\widetilde{Y}\) to solve for all the equivalent radiances emitted by the grids that are within the instantaneous footprint to which \(Y_{i}\) corresponds. To identify the corresponding elements within \(\beta\) to be solved in the \(i^{\text{th}}\) local regression, the column vector \(c(i)\) that collects the positions of the elements in \(\beta\) to be included in the \(i^{\text{th}}\) local scope is defined. These elements in the column vector \(c(i)\) represent the positions of the non-zero elements in the \(i^{\text{th}}\) row of \(\matr{X}\). Then, the column vector \(r(i)\) that collects the positions of the at-sensor radiance observations in \(\widetilde{Y}\) to be used in the \(i^{\text{th}}\) local regression is the collection of the positions of the rows in \(\matr{X}\) that contain at least one non-zero element in the columns indicated in the column vector \(c(i)\). With \(c(i)\) and \(r(i)\) that define the \(i^{\text{th}}\) local scope, the \(i^{\text{th}}\) local regression can be expressed as Equation (3).
	
	\begin{equation}
	\widetilde{Y}^{(i)} = \matr{X}^{(i)} \beta^{(i)} + \tilde{\epsilon}^{(i)}
	\end{equation}
	
	where \(\widetilde{Y}^{(i)}\) is a column vector in which the \(j^{\text{th}}\) element is the element in \(\widetilde{Y}\) at the position indicated by the \(j^{\text{th}}\) element of \(r(i)\), and \(\beta^{(i)}\) is a column vector in which the \(j^{\text{th}}\) element is the element in \(\beta\) at the position indicated by the \(j^{\text{th}}\) element of \(c(i)\). \(\matr{X}^{(i)}\) is a matrix with the appropriate size that characterizes the weighted-sum relationship between the at-sensor radiance observations in \(\widetilde{Y}^{(i)}\) and the equivalent radiances emitted by the grids in \(\beta^{(i)}\). However, due to the exclusion of the elements in \(\beta\) via \(c(i)\), directly filling the locations in \(\matr{X}^{(i)}\) with the corresponding elements in \(\matr{X}\) is not appropriate. The scale parameter \(\phi^{(i)}_{j}\) needs to be applied to the corresponding elements in \(\matr{X}\) before assigning these elements to the \(j^{\text{th}}\) row of \(\matr{X}^{(i)}\) to ensure that the sum of the elements in each row of \(\matr{X}^{(i)}\) is one. This sum-to-one adjustment for \(\matr{X}^{(i)}\) is based on the assumption that the biased expressions for the elements in \(\widetilde{Y}^{(i)}\) due to the exclusion of the elements in \(\beta\) via \(c(i)\) can be corrected through scaling the weights for the elements in \(\beta^{(i)}\). However, this scaling approach for the bias correction also amplifies the uncertainty when expressing the elements in \(\widetilde{Y}^{(i)}\) as linear functions of \(\beta^{(i)}\). Therefore, the \(j^{\text{th}}\) element of the column vector \(\tilde{\epsilon}^{(i)}\) is expressed as the scale parameter \(\phi^{(i)}_{j}\) multiplied by the sum of the noise \({\epsilon_{k_{1},k_{2}}}\) associated with the non-zero element \(x_{k_{1},k_{2}}\) in \(\matr{X}\) over all the possible \(k_{2}\) listed in \(c(i)\) where \(k_{1}\) is the position indicated by the \(j^{\text{th}}\) element of \(r(i)\). 
	
	{\it Local estimation}. By adopting the weighted least squares approach, the estimator \(\hat{\beta}^{(i)}\) for estimating \(\beta^{(i)}\) in Equation (3) can be expressed as Equations (4) and (5).
	
	\begin{gather}
	\hat{\beta}^{(i)} = \matr{P}^{(i)} \widetilde{Y}^{(i)} \\
	\intertext{with}
	\matr{P}^{(i)} = ({\matr{X}^{(i)}}^{T} \matr{W}^{(i)} \matr{X}^{(i)})^{-1} {\matr{X}^{(i)}}^{T} \matr{W}^{(i)}
	\end{gather}
	
	where the matrix \(\matr{W}^{(i)}\) is a diagonal matrix in which the \(j^{\text{th}}\) diagonal element is inversely proportional to the variance of the \(j^{\text{th}}\) element of \(\tilde{\epsilon}^{(i)}\). Based on the definition of the \(\tilde{\epsilon}^{(i)}\) vector, the \(j^{\text{th}}\) diagonal element of \(\matr{W}^{(i)}\) can be expressed as the reciprocal of the sum of \((\phi^{(i)}_{j} x_{k_{1},k_{2}})^{2}\) over all the positions \(k_{2}\) listed in \(c(i)\) where \(k_{1}\) is the position indicated by the \(j^{\text{th}}\) element of \(r(i)\). 
	
	Based on Equations (3), (4), and (5), it is clear that the expectation of \(\hat{\beta}^{(i)}\) is equal to \(\beta^{(i)}\). Based on this unbiased property of the estimator \(\hat{\beta}^{(i)}\), the covariance between the \(j^{\text{th}}\) element \(\hat{\beta}^{(i)}_{j}\) in \(\hat{\beta}^{(i)}\) and the \(j'^{\text{th}}\) element \(\hat{\beta}^{(i')}_{j'}\) in \(\hat{\beta}^{(i')}\) can further be derived and expressed as Equations (6) and (7).
	
	\begin{align}
	Cov(\hat{\beta}^{(i)}_{j},\hat{\beta}^{(i')}_{j'}) & = E\left[\matr{P}^{(i)}_{j} \tilde{\epsilon}^{(i)} \matr{P}^{(i')}_{j'} \tilde{\epsilon}^{(i')}\right]  \\
	& = \sigma^{2} \sum_{ \mathclap{ \{(k,k') \mid r(i)_{k}=r(i')_{k'}\} } } \Big( \phi^{(i)}_{k} \phi^{(i')}_{k'} \matr{P}^{(i)}_{j,k} \matr{P}^{(i')}_{j',k'} \sum_{ \mathclap{ \{l \mid l \in c(i) \cap c(i')\} } } x^{2}_{r(i)_{k},l} \Big)
	\end{align}
	
	where \(\matr{P}^{(i)}_{j}\) and \(\matr{P}^{(i')}_{j'}\) represent the row vector of the \(j^{\text{th}}\) row of the projection matrix \(\matr{P}^{(i)}\) and the row vector of the \(j'^{\text{th}}\) row of the projection matrix \(\matr{P}^{(i')}\), respectively. \(\matr{P}^{(i)}_{j,k}\) and  \(\matr{P}^{(i')}_{j',k'}\) represent the element in the \(j^{\text{th}}\) row and \(k^{\text{th}}\) column of the projection matrix \(\matr{P}^{(i)}\) and the element in the \(j'^{\text{th}}\) row and \(k'^{\text{th}}\) column of the projection matrix \(\matr{P}^{(i')}\), respectively. \(r(i)_{k}\) and \(r(i')_{k'}\) are the \(k^{\text{th}}\) element of the position vector \(r(i)\) and the \(k'^{\text{th}}\) element of the position vector \(r(i')\), respectively. To simplify the notations, the \(c(i)\) and \(c(i')\) in Equation (7) represent the set that collects all the position elements in the previously defined vector \(c(i)\) and the set that collects all the position elements in the previously defined vector \(c(i')\), respectively. Because the elements in \(\tilde{\epsilon}^{(i)}\) and \(\tilde{\epsilon}^{(i')}\) can be expressed as the sums of the mutually independent noises associated with the non-zero elements in the matrix \(\matr{X}\), Equation (6) can be rewritten as the sum of the expectations of the products of the pairs of these noises with zero mean. Due to these independent and zero-mean properties, Equation (7) can be derived by excluding the expectations that return zero in the summation.
	
	{\it Global estimation}. For each local regression, a subset of the elements in \(\beta\) is estimated. In other words, each element in \(\beta\) can be estimated via different local regressions. Based on the weighted least squares estimation approach, these estimations that come from different local regressions for an element in \(\beta\) are all unbiased estimators with different dependencies between each other as described by Equations (6) and (7). To fully utilize all the information that comes from different local regressions, the global estimation for each element in \(\beta\) should be a linear combination of all the local estimators for that element by using the dependencies between the corresponding local estimators as the reference. Therefore, the global estimator \(\hat{\beta}_{j}\) for the \(j^{\text{th}}\) element in \(\beta\) can be expressed as Equation (8).
	
	\begin{equation}
	\hat{\beta}_{j} = \frac{ {\eta^{(j)}}^{T} }{ {\mathbf{1}}^{T} {\eta^{(j)}} } \gamma^{(j)} 
	\end{equation}
	
	where \(\gamma^{(j)}\) is a column vector that collects the \(k^{\text{th}}\) element of \(\hat{\beta}^{(i)}\) for all \(i\) and \(k\) that satisfy the relationship that the \(k^{\text{th}}\) element of the position vector \(c(i)\) is \(j\). The column vector \(\eta^{(j)}\) represents the eigenvector to which the largest eigenvalue of the correlation matrix of the column vector \(\gamma^{(j)}\) calculated from Equations (6) and (7) corresponds. \(\mathbf{1}\) is a column vector, in which all the elements are one, with the same length as the vector \(\eta^{(j)}\) to ensure that \(\hat{\beta}_{j}\) is an unbiased estimator of the \(j^{\text{th}}\) element in \(\beta\).
	
	\section{Numerical Simulations}
	{\it Stochastically simulated artificial image.} To evaluate the performance of the proposed method in retrieving the information that comes from oversampling, artificial images that are constituted by pixels with stochastically simulated values are adopted as the image that displays the true equivalent radiances emitted by the grids defined on the Earth surface. To create textures in these artificial images to simulate ground objects, the pixel values in each artificial image are simulated sequentially from the top left corner to the bottom right corner. The first pixel value is simulated from the standard Gaussian distribution. The value of the subsequent pixel is simulated from the Gaussian distribution that is conditional on all the pixel value(s) that has(have) been simulated within a radius of \(R\) pixels from the current pixel, with the assumption that all the pixel values follow the standard Gaussian distribution and jointly follow the multivariate Gaussian distribution. The required correlation(s) between pixel values for calculating the mean and the variance of this conditional Gaussian distribution is(are) evaluated via the spherical variogram model \(V(.)\) indicated in Equation (9).
	
	\begin{equation}
	V(d) = \left( \frac{3d}{2R} - \frac{d^{3}}{2R^{3}} \right) 1_{(0,R)}(d) + 1_{[R,\infty)}(d)
	\end{equation}	

	where \(d\) represents the Euclidean distance in pixel units between two pixels in the images. To control the image histogram, each pixel value in an artificial image simulated based on the Gaussian assumption is converted to the corresponding standard Gaussian percentile via the cumulative distribution function of the standard Gaussian distribution. This calculated percentile is then converted into the corresponding Gamma quantile, via the inverse of the Gamma cumulative distribution function with a fixed shape parameter \(\alpha\) and a fixed scale parameter \(\lambda\) that are adopted to all the pixels in this artificial image, as the final pixel value. As an illustrative example, a simulated artificial image is provided in Figure 1.

	\begin{figure}[h!]
		\includegraphics[width=\linewidth]{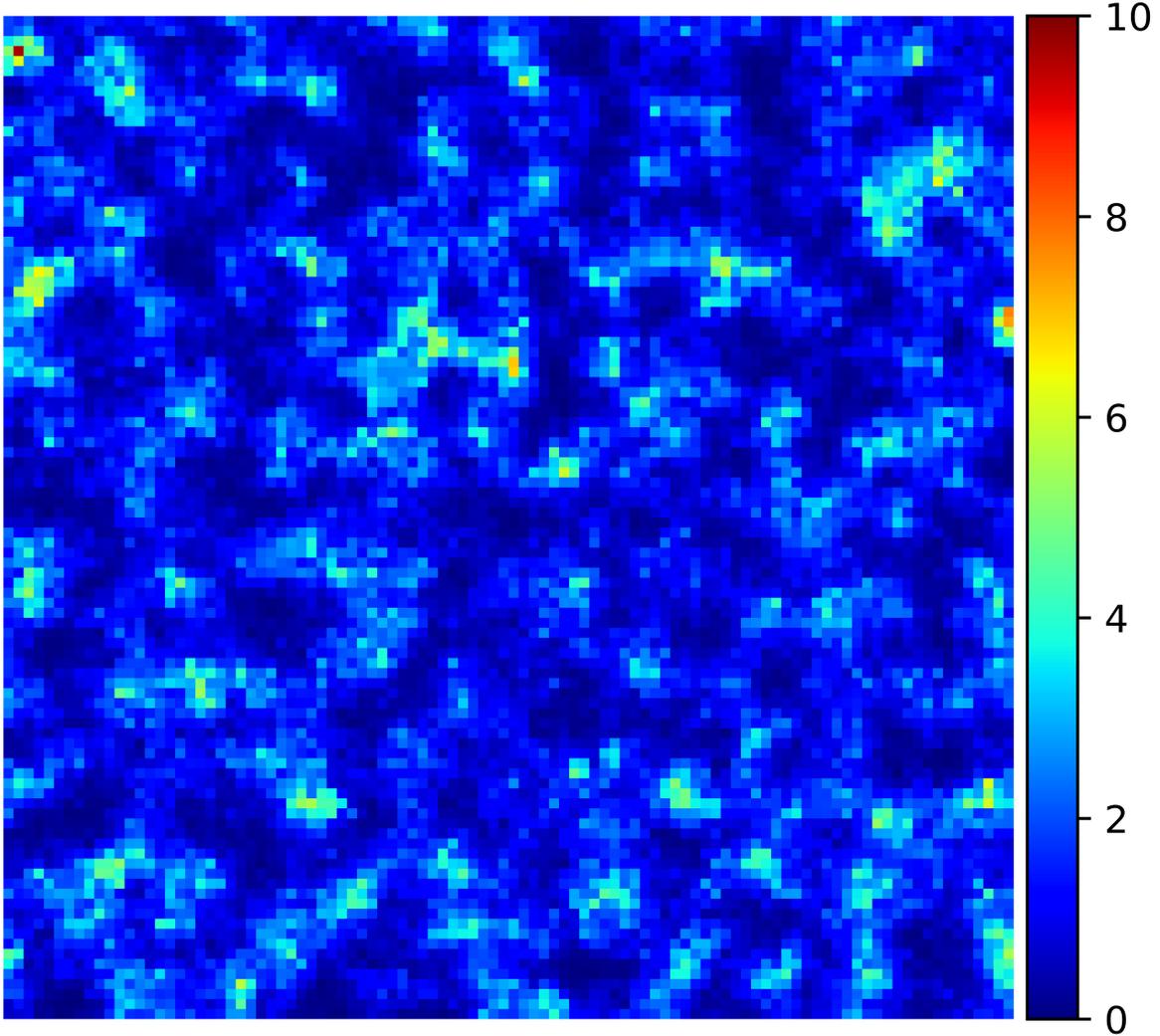}
		\caption{An illustrative example of the stochastically simulated artificial images. The radius \(R\), the shape parameter \(\alpha\), and the scale parameter \(\lambda\) are \(6\) pixels, \(\frac{16}{9}\), and \(\frac{3}{4}\), respectively.}
		\label{Figure 1}
	\end{figure}
	
	{\it Spatially oversampled image.} To create spatially oversampled images for evaluating the performance of the proposed approach, one of the three resampling schemes illustrated in Equations (10), (11), and (12) is applied to the simulated artificial images, with stochastically simulated noises attached. For each pixel in an oversampled image, the weighted average based on the selected resampling scheme is obtained by first matching the center element of the selected resampling matrix to the same pixel location in the corresponding artificial image. Then, via applying each weight specified in the selected resampling matrix to the pixel value in the artificial image at the corresponding location that is relative to the center of the matrix, this weighted average value can be calculated. To simulate and attach the noise to this weighted average as described in Equation (2), a value is independently simulated for each element in the selected resampling matrix, from the Gaussian distribution with zero mean and the standard deviation that is equal to the value of this element multiplied by a fixed constant \(\ddot{\sigma}\) that is adopted to all the pixels in this oversampled image. Then, the sum of all these independently simulated values and this weighted average value is assigned to the corresponding pixel of this oversampled image as the final pixel value. Furthermore, since the noise level of an oversampled image is controlled by the constant \(\ddot{\sigma}\), the ratio of the variance \(\alpha\lambda^{2}\) of the Gamma distribution used for generating the corresponding artificial image to \(\ddot{\sigma}^{2}\) is defined as the signal-to-noise ratio of the oversampled image in this study. As an illustrative example, the corresponding oversampled image obtained from the artificial image demonstrated in Figure 1 is provided in Figure 2.

	\begin{equation}
		\begin{bmatrix}
			0.0267 & 0.1489 & 0.0267 \\
			0.1489 & 0.2976 & 0.1489 \\
			0.0267 & 0.1489 & 0.0267
		\end{bmatrix}
	\end{equation}
	
	where the value of each element is proportional to the cosine of the product of \(\frac{\pi}{3}\) and the Euclidean distance from that element to the center of the matrix.
	
	\begin{equation}	
		\begin{bmatrix}
			0.0069 & 0.0302 & 0.0388 & 0.0302 & 0.0069 \\
			0.0302 & 0.0573 & 0.0672 & 0.0573 & 0.0302 \\
			0.0388 & 0.0672 & 0.0776 & 0.0672 & 0.0388 \\
			0.0302 & 0.0573 & 0.0672 & 0.0573 & 0.0302 \\
			0.0069 & 0.0302 & 0.0388 & 0.0302 & 0.0069 
		\end{bmatrix}
	\end{equation}
	
	where the value of each element is proportional to the cosine of the product of \(\frac{\pi}{6}\) and the Euclidean distance from that element to the center of the matrix.
		
	\begin{equation}
		\begin{bmatrix}
			0.0032 & 0.0111 & 0.0163 & 0.0181 & 0.0163 & 0.0111 & 0.0032 \\
			0.0111 & 0.0199 & 0.0257 & 0.0277 & 0.0257 & 0.0199 & 0.0111 \\
			0.0163 & 0.0257 & 0.0318 & 0.0340 & 0.0318 & 0.0257 & 0.0163 \\
			0.0181 & 0.0277 & 0.0340 & 0.0364 & 0.0340 & 0.0277 & 0.0181 \\
			0.0163 & 0.0257 & 0.0318 & 0.0340 & 0.0318 & 0.0257 & 0.0163 \\
			0.0111 & 0.0199 & 0.0257 & 0.0277 & 0.0257 & 0.0199 & 0.0111 \\
			0.0032 & 0.0111 & 0.0163 & 0.0181 & 0.0163 & 0.0111 & 0.0032 
		\end{bmatrix}		
	\end{equation}
	
	where the value of each element is proportional to the cosine of the product of \(\frac{\pi}{9}\) and the Euclidean distance from that element to the center of the matrix.

	\begin{figure}[h!]
	\includegraphics[width=\linewidth]{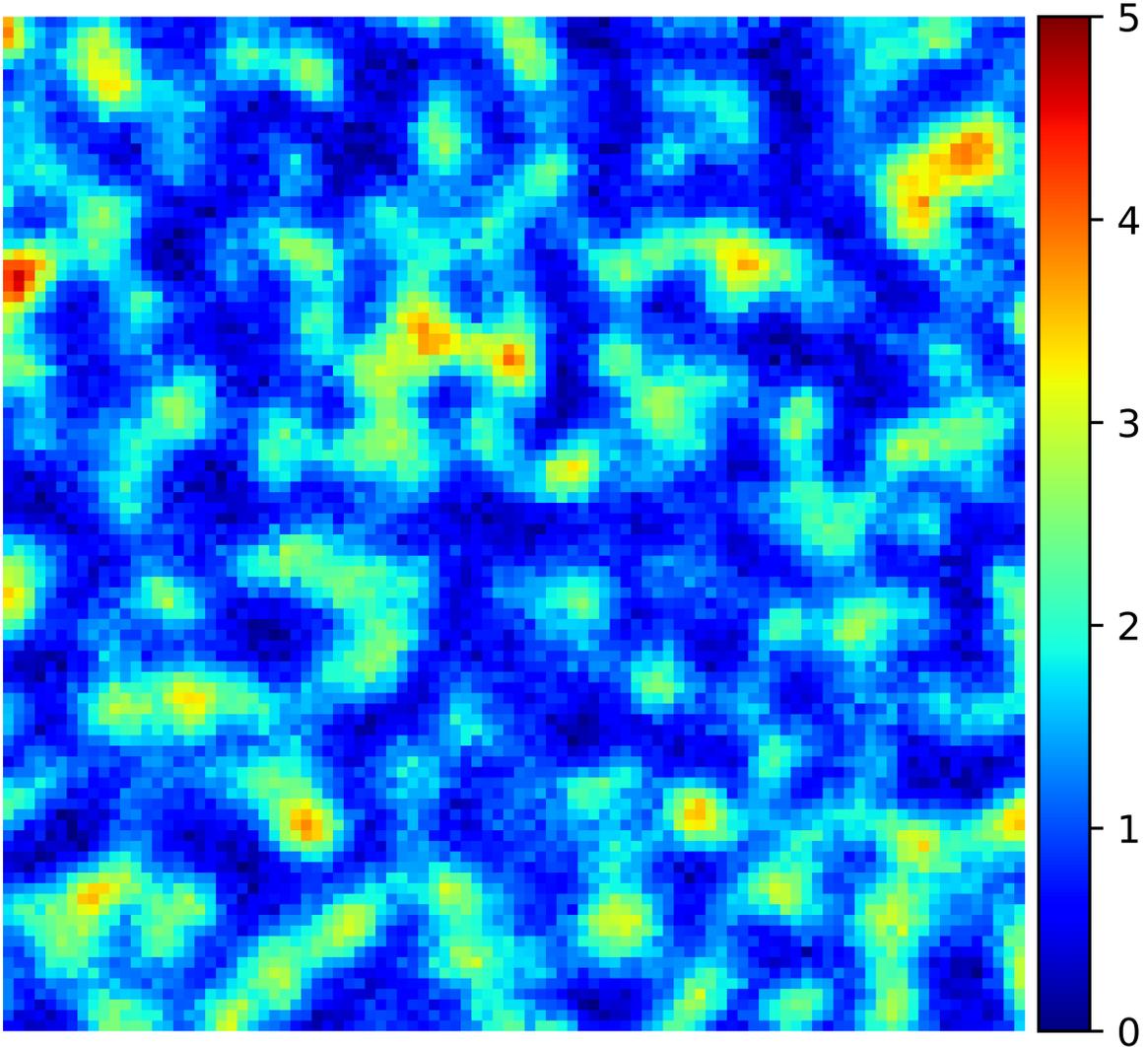}
	\caption{An illustrative example of the spatially oversampled images. Equation (11) is adopted for the resampling. \(\ddot{\sigma}^{2}\) is set to \(\frac{1}{2}\) for the simulated noises. Two pixels are missing from each side of the image due to the application of Equation (11).}
	\label{Figure 2}
	\end{figure}

	{\it Reconstructed image.} By rearranging the pixels in an oversampled image and the pixels in the corresponding artificial image to form the column vector \(\widetilde{Y}\) and the column vector \(\beta\) indicated in Equation (2), respectively, the matrix \(\matr{X}\) can be obtained accordingly based on the weighted average calculation procedure that adopts one of the three resampling schemes indicated in Equations (10), (11), and (12) to obtain the oversampled image. Then, Equations (4) and (5) can be adopted to calculate the local estimator \(\hat{\beta}^{(i)}\) for each row \(i\) in \(\widetilde{Y}\). With the calculated local estimators, the global estimator \(\hat{\beta}_{j}\) for each row \(j\) in \(\beta\) can be evaluated via Equations (7) and (8). By assigning the value of each evaluated global estimator to the corresponding pixel in the artificial image, the reconstructed image can be obtained. As an illustrative example, the image reconstructed from the oversampled image shown in Figure 2 is provided in Figure 3.

	\begin{figure}[h!]
	\includegraphics[width=\linewidth]{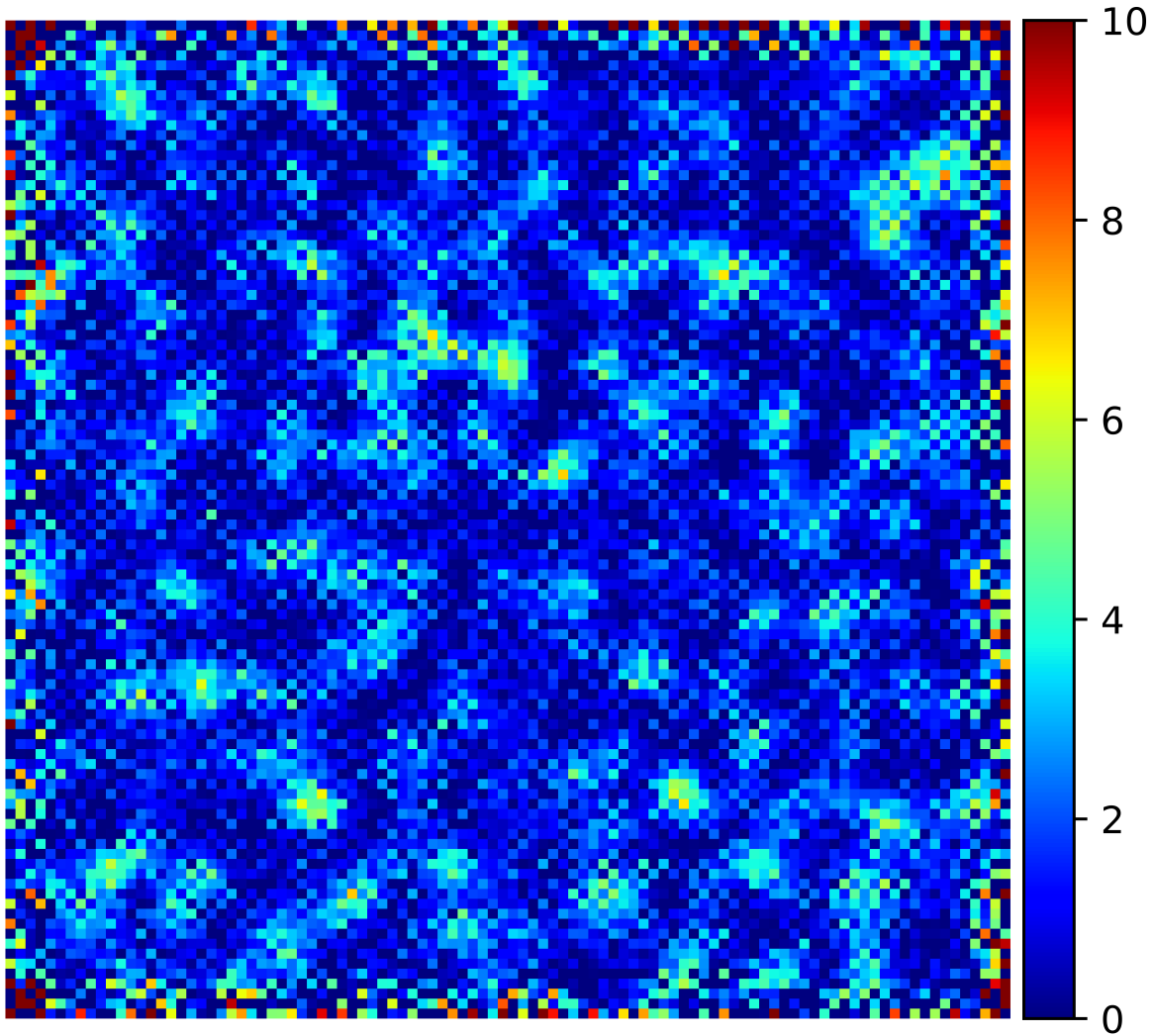}
	\caption{An illustrative example of the reconstructed images.}
	\label{Figure 3}
	\end{figure}
	
	As demonstrated in Figure 3, the application of the proposed method is able to recover the texture of the artificial image illustrated in Figure 1 from the oversampled image shown in Figure 2. In addition, the noises that are simulated and attached to the oversampled image for mimicking the sensor noises are spatially randomly distributed by the proposed approach in the reconstructed image. The collection of the pixel-wise differences between the reconstructed image demonstrated in Figure 3 and the artificial image illustrated in Figure 1 further indicates that these spatially-randomly-distributed noises are evenly distributed around zero as well. This capability of recovering the texture and this capability of evenly distributing the noises suggest the effectiveness of the proposed method in improving the spatial resolution of an image via oversampling. 
	
	{\it Feasibility.} To further evaluate the feasibility of the proposed approach in different scenarios, different values of \(\alpha\) and \(R\) are adopted to generate artificial images with different textures. Oversampled images with different degrees of oversampling are created via Equations (10), (11), and (12). Different values of \(\lambda\) and \(\ddot{\sigma}\) are applied to reproduce different signal-to-noise ratios. For each setting of the combinations of the textures, the degrees of oversampling, and the signal-to-noise ratios, 100 artificial images are simulated with a length of 40 pixels and a width of 40 pixels. For each simulated artificial image, the corresponding oversampled image is created via applying the corresponding resampling matrix and simulating noises based on the corresponding value of \(\ddot{\sigma}\) to attach. Then, the mean, the standard deviation, and the skewness of the pixel-wise differences between the reconstructed image obtained via applying the proposed method to this oversampled image and the artificial image adopted for creating this oversampled image, are calculated. The three averages of these three metrics calculated from the 100 sets of the collected pixel-wise differences then summarize the performance of the proposed approach for this particular combination of the textures, the degrees of oversampling, and the signal-to-noise ratios. By collecting the comparison results from all the scenarios examined in this study, Table 1 is formed to summarize the feasibility of the proposed method.
	
	\begin{table}[h!]
		\begin{center}
	\includegraphics[width=0.9\linewidth]{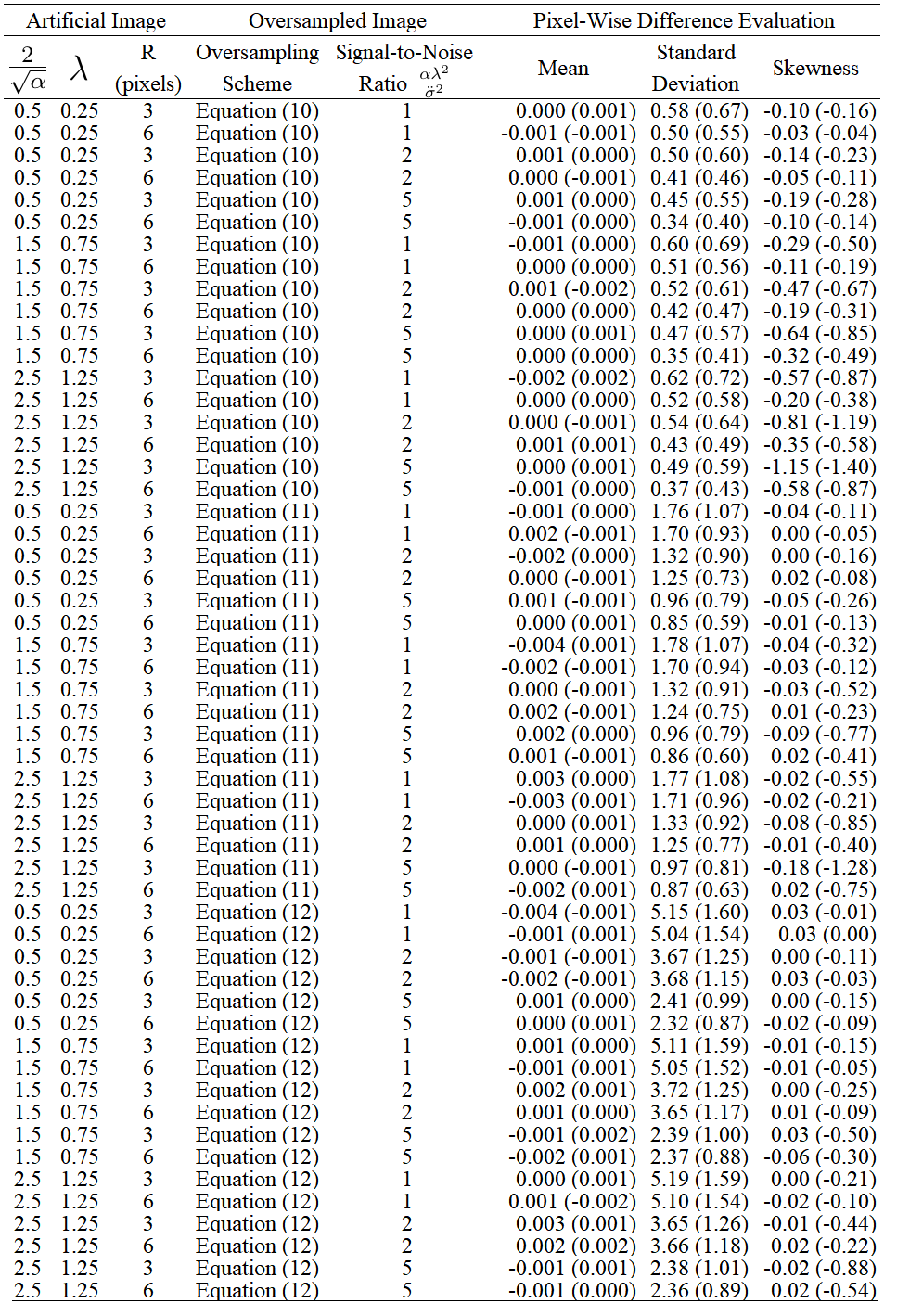}
	\caption{The summary of the feasibility of the proposed method. The pixel-wise differences are obtained via subtracting the pixel values in the artificial images from the corresponding pixel values in the corresponding reconstructed images. Two, three, and four pixels from each side of the images are excluded from the pixel-wise-difference calculations when the size of the resampling matrix is three-by-three, five-by-five, and seven-by-seven, respectively. As a comparative base, the values in parentheses are the corresponding results when the elements in Equations (10), (11), and (12) are replaced by the values: \(\frac{1}{9}\), \(\frac{1}{25}\), and \(\frac{1}{49}\), respectively.}
	\label{Table 1}
		\end{center}
	\end{table}	

	With the feasibility summary described in Table 1, the pixel-wise performance evaluation of the parameter setting adopted to generate Figure 1, Figure 2, and Figure 3 can be identified. This identified performance evaluation information can then be used as a comparative base for interpreting Table 1. With this comparative base, whether or not the proposed method is able to recover the textures in the artificial images from the oversampled images, should be evaluated by using, whether or not the mean and skewness metrics of the pixel-wise differences are close to zero, as the criterion. Then, the standard deviation metric of the pixel-wise differences simply reflects the level of the noises attached to the reconstructed images. As illustrated in Table 1, the skewness metric of the pixel-wise differences decreases when \(\frac{2}{\sqrt{\alpha}}\) increases, \(R\) decreases, or the size of the resampling matrix decreases. This decrease in the skewness metric implies that the proposed approach relies on, obtaining an appropriate expression for at-sensor radiance observations by using the radiances emitted by fractions of the corresponding instantaneous footprints of the sensor, to enhance the spatial resolution of oversampled images. When making inferences on the pixel values in an image to be reconstructed by using the adjacent pixels is not appropriate, due to the high image skewness of the image to be reconstructed, the low spatial dependence within the image to be reconstructed, or the low overlap rate of the applied oversampling scheme, the application of the proposed method without an additional correction to the distortion of the image histogram of the reconstructed image may not be appropriate. Furthermore, as demonstrated in Table 1, the standard deviation metric of the pixel-wise differences increases when \(\frac{\alpha\lambda^{2}}{\ddot{\sigma}^{2}}\) decreases, or the size of the resampling matrix increases. This increase in the standard deviation metric indicates that, to effectively adopt the proposed approach to retrieve information via oversampling, arranging an appropriate oversampling scheme is essential when designing a sensor to ensure that the noises attached to the reconstructed images are controlled at a reasonable level.
	
	\section{Conclusions}
	To enhance the spatial resolution of oversampled images, the concept of the regression decomposition and synthesis is proposed. With the concept of the regression decomposition, the spatial resolution can be enhanced by using the segments of an oversampled image as units, to improve the reliability of the enhancement result of each segment. With the concept of the regression synthesis, the enhancement results of these segments can be integrated via the statistical inferences on the estimation variances to ensure the consistency of the spatial resolution enhancement throughout the entire image to further reinforce the reliability of the enhancement result.
	
	To evaluate the performance of the proposed method, oversampled images are generated via applying oversampling schemes to stochastically simulated images, with stochastically simulated noises attached. The implementation of the proposed approach to these generated oversampled images demonstrates the effectiveness of the proposed method in enhancing the spatial resolution, when comparing the resulting images with the corresponding stochastically simulated images. Based on different combinations of the simulated image textures, the degrees of oversampling, and the signal-to-noise ratios, these comparison results further conclude the feasibility of proposed method in practical applications.

\end{document}